# In vivo endoscopic autofluorescence microspectro-imaging of bronchi and alveoli

G. Bourg-Heckly[*a], L. Thiberville[b], C. Vever-Bizet[a], B. Viellerobe[c]
[a]BioMoCeTi Laboratory, CNRS UMR 7033, Université Pierre et Marie Curie Paris 6, France
[a]Clinique Pneumologique, Rouen University Hospital, Rouen, France
[a]Mauna Kea Technologies, Paris, France

## ABSTRACT

Fibered confocal fluorescence microscopy (FCFM) is a new technique that can be used during a bronchoscopy to analyze the nature of the human bronchial and alveolar mucosa fluorescence microstructure. An endoscopic fibered confocal fluorescence microscopy system with spectroscopic analysis capability was developed allowing real-time, simultaneous images and emission spectra acquisition at 488 nm excitation using a flexible miniprobe that could be introduced into small airways. This flexible 1.4 mm miniprobe can be introduced into the working channel of a flexible endoscope and gently advanced through the bronchial tree to the alveoli. FCFM in conjunction with bronchoscopy is able to image the *in vivo* autofluorescence structure of the bronchial mucosae but also the alveolar respiratory network outside of the usual field of view. Microscopic and spectral analysis showed that the signal mainly originates from the elastin component of the bronchial subepithelial layer. In non smokers, the system images the elastin backbone of the aveoli. In active smokers, a strong autofluorescence signal appears from alveolar macrophages. The FCFM technique appears promising for *in vivo* exploration of the bronchial and alveolar extracellular matrix.

## 1. INTRODUCTION

Autofluorescence bronchoscopy has been extensively avaluated during the past decade[1]: fluorescences bronchoscopy is based on the observation that premalignant and malignant bronchial mucosae fluoresce less than normal tissue. This allows detection of lesions that may have a normal appearance during conventional white-light bronchoscopy[2]. However, the technique is hampered by the low specificity of the fluorescence defect, which ranges from 25 to 50%[1].

Coupled with autofluorescence bronchoscopy, the use of a method that would allow real-time non invasive histologic imaging—a principle that is also referred to as "optical biopsy"—may help to ensure higher yield biopsy samples, increase the specificity of the endoscopic technique, and potentially avoid unnecessary biopsy sampling or repeated procedures.

Until now, the external diameter of the thinnest commercially available bronchoscope does not make it possible to image the respiratory tract beyond bronchioles smaller than 3mm in diameter in-vivo[4]. Therefore, the pathology of the terminal respiratory unit, from the small terminal and respiratory bronchioles down to the alveolar acini, was until recently assessed only in vitro, using techniques such as bronchoalveolar lavage and histology from transbronchial or open lung biopsies. No real time imaging was available.

Fibered confocal microscopy is a new technique that can be used to image the microscopic structure of a living tissue[3]. Fibered confocal microscopy is based on the principle of confocal microscopy, which provides a clear, in-focus image of a thin section within a biological sample, where the microscope objective is replaced by a flexible fiberoptic miniprobe. In its fluorescence mode (fibered confocal fluorescence microscopy [FCFM]), the technique makes it possible to obtain high-quality images from endogenous or exogenous tissue fluorophores, through a fiberoptic probe of 1.4 mm diameter or less that can be introduced into the 2 mm working channel of a flexible bronchoscope.

Human connective tissue of the peripheral lung is made of about 50% elastin in humans (5, 6). We therefore hypothesize that FCFM in conjunction with bronchoscopy is able to assess the microstructure of the alveolar elastin network in vivo.

In this study, we use FCFM to obtain real time in-vivo imaging of the respiratory bronchi and alveoli in humans during bronchoscopy: we describe the *in vivo* autofluorescence microscopic structure of the normal bronchial mucosae and the alterations of the bronchial microstructure in precancerous lesions and we characterize the confocal microanatomy in alveoli in both smoker and non smoker healthy subjects.

## 2. METHODOLOGY

**2.1 Fibered Confocal Fluorescence imaging and spectroscopic system**

The FCM imaging instrument, the F400S, is based on the fibered confocal fluorescence microscopy technology system developed by Mauna Kea Technology (Paris, France) : it is basically identical to the commercially available FCFM device called "Cellvizio® -Lung (Mauna Kea Technologies) except that a spectroscopic channel to the device allows simultaneous recording of the spectrum and the microscopic images of the observed field of view.

As can be seen on figure 1, this device consists of four main elements : a Laser Scanning Unit, a miniaturized flexible fibered probe, a spectroscopic channel and a dedicated control and acquisition software. The laser scanning unit is composed mainly of a solid state laser diode (Laser Sapphire, Coherent, Santa Clara, CA, USA) emitting at 488 nm, an XY scanning system injecting the laser beam into the fiber bundle one fiber at a time and a photodetector. Each fiber, of 1.9 µm core diameter, acts as a point source and a point detector, thereby ensuring the confocal properties of the system. The bundle is scanned at high speed resulting in an frame rate of 12 images per second for a 896 X 640 pixels image. The illumination of the tissue is thus assured in turn by only one fiber which, in return, collects the fluorescence light. On return, the optical path is divided into 2 parts by a beamsplitter, 80% of the fluorescence signal being used for imaging and 20% for spectral analysis. The processed images are displayed on the monitor in real-time. The fluorescence signal, for image and spectral analysis as well, is recorded between 500 nm and 750 nm.

**2.2 Probes**
For the studies reported here, the microprobe is a Alveoflex Confocal Miniprobe® (Mauna Kea Technologies) with an overall diameter of 1.4 mm compatible with the operating channel of standard bronchoscopes for *in vivo* proximal and distal lung tissues exploration. The fiber bundle is made of 30,000 fiber cores. This probe produces images from a layer of 0 to 50 µm in depth below surface with a lateral resolution of 3.5 µm. Because the probe is devoid of distal optics, the field of view used in the study strictly aquals the scanned surface (600 µm).

**2.3 Data processing**

*Images*
Raw data images are processed by a dedicated image processing software developed in-house by MKT (Medicell) which removes the fiber bundle pattern and provides sequences of images in real time together with various quantification tools (distance measurements and statistics).

*Spectra*
The fiber background spectrum was substracted from all raw autofluorescence tissular spectra. Then the spectra were corrected for the spectral sensitivity of the spectrometer detector and optical setup. For comparison of spectral distributions, spectra were normalized with respect to their maximum intensity.

**Reference spectra**

In order to attempt to identify the endogenous fluorophores most likely involved in the autofluorescence image formation, pure collagen and elastin emission reference spectra were acquired with the F400S instrument under the same experimental conditions as used for endoscopic measurements. Collagen 1 gels were kindly provided by Gervaise Mosser from « Chimie de la Matière Condensée » laboratory (université Pierre et Marie Curie, Paris) ; they were obtained à compléter. Elastin from human lung prepared by non degradative extraction was purchased fronm Sigma (reference E7152).

## 3. RESULTS

Fig. 1. Schematic diagram of the prototype F400/S (Mauna Kea Technologies), a dual fibered confocal imaging and spectroscopic platform.

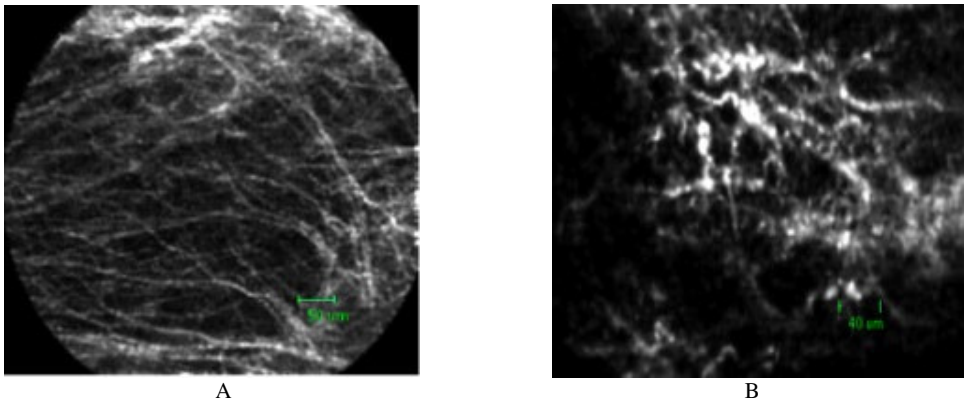

Fig. 2. *in vivo* in human bronchi, subepithelial connective tissue layer imaging: A, origin of the lingual. B, carcinoma in situ

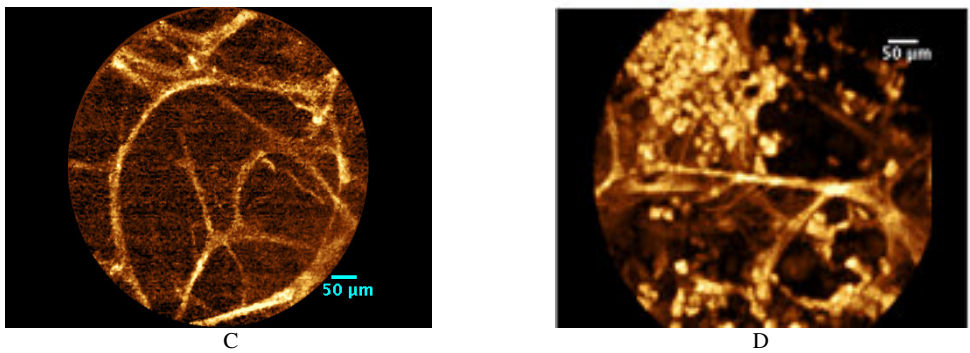

Fig. 3. *in vivo* FCFM alveolar imaging : in non smoker, C ; in healthy smoker volunteer, D